\documentclass[a4paper,fleqn,usenatbib]{mnras}

\usepackage{newtxtext,newtxmath}
\usepackage[T1]{fontenc}
\usepackage{ae,aecompl}

\usepackage{bm}
\usepackage{graphicx}
\usepackage{amsmath}	
\usepackage{amssymb}	
\usepackage{color}

\def\au{\,\mathrm{a.u.}}

\def\Ry{\,\mathrm{Ry}}
\def\kelvin{\,\mathrm{K}}

\def\angstrom{\textrm{\AA}}

\title[PI of Sr, Y$^{+}$, Zr$^{2+}$]
      {Opacity modelling of heavy-metal hot subdwarfs.
       Photoionization of Sr$^{0}$, Y$^{+}$ and Zr$^{2+}$.}
\author[L. Fern\'{a}ndez-Menchero et al.]{
        L. Fern\'{a}ndez-Menchero$^{1}$\thanks{E-mail: l.fernandezmenchero@qub.ac.uk (LFM)}, 
        C. S. Jeffery$^{2}$, 
        C. A. Ramsbottom$^{1}$ and 
        C. P. Ballance$^{1}$.
        \\
        $^{1}$Centre of Theoretical Atomic, Molecular and Optical Physics.
        Queen's University Belfast.
        University Road, Belfast BT7 1NN. \\
        $^{2}$Armagh Observatory, College Hill, Armagh BT61 9DG.}
\date{Accepted XXX. Received YYY; in original form ZZZ}
\pubyear{2019}

\begin{document}
\label{firstpage}
\pagerange{\pageref{firstpage}--\pageref{lastpage}}
\maketitle

\begin{abstract}

Heavy-metal hot subdwarfs (sdB and sdO) represent a small group of stars 
with unusually high concentrations of trans-iron elements in their atmospheres, 
having abundances $\sim 10 000$ times solar. 
One example is \mbox{LS\,IV-14\degr 116}, where a number
of heavy-metal absorption lines of \ion{Sr}{ii}, \ion{Y}{iii} 
and \ion{Zr}{iv} have been observed in the optical 
band $4\,000 - 5\,000\,\angstrom$.
We use a fully relativistic Dirac atomic R-Matrix (DARC) 
to calculate photoionization cross sections of Sr$^{0}$, Y$^{+}$ and Zr$^{2+}$ 
from their ground state to the twentieth excited level.
We use the cross sections and the oscillator strengths 
to simulate the spectrum of a hot subdwarf.
We obtain complete sets of photoionization cross sections for the three ions 
under study. 
We use these data to calculate the opacity of the stellar atmospheres of hot 
subdwarf stars, and show that for overabundances observed in some heavy-metal 
subdwarves, photo-excitation from zirconium, in particular, does contribute some 
back warming in the model.

\end{abstract}


\section{Introduction}
\label{sec:introduction}

Heavy-metal subdwarfs represent a small fraction of the hot subluminous stars, 
O and B stars lying below the main sequence  \citep{heber2009}. 
The cooler subdwarf B (sdB) stars are low-mass stars with a core of burning helium, 
and a thinner hydrogen envelope.  
The majority have masses close to $0.5{\rm M_{\odot}}$ and are frequently referred to 
as `extreme horizontal-branch stars'. 
Owing to competition between gravitational settling and radiative levitation,  
their photospheres  are  helium poor and relatively metal-rich \citep{geier2013}. 
The hotter subdwarf O stars (sdO) are  similar but hotter and, with no hydrogen envelope, 
the majority lie on or close to the helium main sequence.  
Their surfaces are mostly hydrogen poor. 
In between lie a small number of 'not intermediate helium subdwarfs', 
having surface helium-to-hydrogen ratios between $10$ and $95\%$ 
by number \citep{naslim2012} and effective temperatures ($T_{\rm eff}$) 
between $30\,000$ and $40\,000 \kelvin$.   
Of these, an increasingly large fraction have been discovered to show 
super-abundances of trans-iron elements forming two groups, 
the cooler zirconium-rich and the hotter lead-rich {\it heavy-metal subdwarfs} 
\citep{naslim2011, naslim2013, jeffery2017a, dorsch2019, latour2019, jeffery2019, naslim2020}. 

It is well known that under appropriate conditions, 
radiative levitation can lead to thin layers containing very high concentration 
of specific ions of  heavy atoms in the stellar photosphere
\citep{shulyak2010}.
When this concentration occurs at the precise depth at which absorption lines form,  
abundance measurements return very high values. 
The line formation depth is closely correlated, for any given ion, with $T_{\rm eff}$, 
so that in the range of current interest and, 
specifically in the class prototype, \mbox{LS\,IV-14\degr 116}, the heavy-metal 
ions \ion{Sr}{ii}, \ion{Y}{iii} and \ion{Zr}{iv}, amongst others, were observed with 
abundances $\approx 10\,000$ times solar \citep{naslim2011}.
These abundances are so high that this was the first time some of these ions had been 
seen in stellar  spectra at optical wavelengths,
and required the calculation of additional atomic data. 

Unfortunately, heavy metals (above Kr) are a largely unexplored field in atomic 
processes, with the only exceptions being tungsten and molybdenum for their 
applications in fusion research.
Accurate atomic data for the observed lines of heavy elements in dwarf and 
subdwarf stars are absent in the principal databases.
CHIANTI \citep{delzanna2015c} contains up-to-date data only for elements up to Zn,
OPEN-ADAS\footnote{\url{http://open.adas.ac.uk}} includes data of heavy ions only those
which are interesting for fusion research, for example tungsten,
and the opacity project TOP \citep{cunto1992,cunto1993,opacity1995} 
includes only ions up to iron.
In the NIST data base \citep{nist2018}, data are only present for 
level energies, transition wave lengths and ionization potentials for ions of 
Sr, Y or Zr, and not for transition probabilities, 
such as line strengths $S$, oscillator strengths $f$, $gf$, 
or spontaneous emission $A$.
Einstein $A$-values for transition probabilities for a small set of transitions in the 
ultraviolet can be found in the \citet{kurucz} data base for 
Sr, Y and Zr neutral and once ionized.
Radiative and collisional data for these heavy ions are necessary for the 
correct interpretation of the spectra of metal-rich astrophysical objects.
Specifically, data for photoionization and photoexcitation are necessary to 
construct good models of opacity and atomic diffusion in stellar atmospheres.
Data for electron-impact excitation are also necessary in the case of collisional 
plasmas, in order to construct collisional-radiative models.

Other spectral lines found in these types of stars correspond to the iron-peak elements, 
\ion{Fe}{iii}, \ion{Fe}{iv}, \ion{Ni}{iv}; elements above iron
\ion{Ga}{iv}, \ion{Ge}{iv}, \ion{Ge}{iii}; and the heavy metal \ion{Pb}{iv}.
Studies of iron-peak elements have been carried out recently in electron-impact
excitation by \citet{smyth2019a}, 
and photoionisation by \citet{smyth2018b} and \citet{fernandez-menchero2019a}.
Heavier metals, such as lead, are scarcely explored.

Photoionization of neutral strontium initially in its ground state has been addressed
both experimentally and theoretically.
Experimentally, \citet{hudson1969} measured absolute photoionization cross sections 
for photons with wavelengths between $1646$ and $2028 \angstrom$, 
and \citet{brown1983} between $1400$ and $1900 \angstrom$.
Theoretical studies of photoexcitation were carried out by \citet{aymar1987a} 
and photoionization by \citet{aymar1989} using the R-matrix method and the
multichannel quantum defect theory (MQDT).
More recently, \citet{madine2005} studied the one- and two-photon ionization
of Sr using a time-dependent R-matrix theory in terms of the Floquet approach.
They studied the photon-energy ranges from $0.21$ to $0.27 \au$ and
from $1.105$ to $1.158 \au$ and obtained good agreement with the experiments.

For this paper, we calculated the photoionization of neutral $\mathrm{Sr}$
and its isoelectronic ions $\mathrm{Y}^{+}$ and $\mathrm{Zr}^{2+}$.
We also calculated the radiative transition coefficients between all the 
spectroscopic levels included in the configuration-interaction (CI) 
expansion of their ionization products: 
$\mathrm{Sr}^{+}$, $\mathrm{Y}^{2+}$ and $\mathrm{Zr}^{3+}$. 
For the scattering calculation, we employ a fully relativistic Dirac Atomic R-matrix 
Code (DARC), and to obtain the atomic structure of the target, 
we use the General-purpose Relativistic Atomic Structure Package (GRASP).
For the modeling of the stellar atmospheres of the heavy-metal hot subdwarfs,
we use the public-use program {\sc sterne} \citep{behara2006} 
to obtain the synthetic spectra in a
local thermodynamical equilibrium (LTE) approximation.

The paper is organised as follows,
in section \ref{sec:structure} we describe the atomic structure of the target ions;
in section \ref{sec:scattering} we explain the method used to calculate the 
photoionization cross sections;
in \ref{sec:results} we discuss the most important results for the studied ions;
in section \ref{sec:modelling} we use the calculated data to simulate the
spectra of atmospheres of metal-rich hot subdwarf stars in the optical;
and finally in section \ref{sec:conclusions} we summarise the most important
conclusions of our work.
For a more detailed examination, 
the complete set of results is provided in the online material.
Atomic units are used unless otherwise stated.

\section{Atomic structure}
\label{sec:structure}

To calculate the best quality atomic structure of the target, we use the
General-purpose Relativistic Atomic Structure Package 
(GRASP) \citep{dyall1989,parpia1996}.
GRASP uses a multi-configuration Dirac-Fock (MCDF) method to calculate the 
fully relativistic wave functions of all the atomic levels of the atom or ion 
in a Coulomb-Dirac framework.
The relativistic multi-electron wave function is expanded in terms of a 
configuration interaction (CI) basis set.
The diagonalization of the Dirac-Fock Hamiltonian is carried out through an
iterative procedure, until the convergence of the atomic orbitals is reached.
One of the problems of the MCDF method concerns convergence, 
which is often very difficult to achieve.
This convergence worsens for lowly charged ions, precisely the ones under 
investigation in this paper \citep{froese-fischer2007}.

We use the same set of configurations in the CI expansion of the three
$\mathrm{Rb}$-like ions.
We include a total of 34 configurations, including single, double and triple 
excitations from the ground configurations,
$\mathrm{4p^6\,5s}$ for $\mathrm{Sr}^{+}$ and $\mathrm{Y}^{2+}$, 
and $\mathrm{4p^6\,4d}$ for $\mathrm{Zr}^{2+}$.
For a proper description of the photoionization process initially in excited levels,
we also have to include the principal ionization channels of these lower excited 
configurations of the initial ions:
$\mathrm{4d^2}$, $\mathrm{5s^2}$, $\mathrm{4d\,5s}$, $\mathrm{4d\,5p}$, 
$\mathrm{5s\,5p}$, $\mathrm{4d\,6s}$, $\mathrm{5s\,6s}$, $\mathrm{5s\,6p}$,
$\mathrm{5p^2}$, $\mathrm{5s\,4f}$, $\mathrm{5s\,5d}$
(see NIST data tables \citep{nist2018}).
We have into take in account in the basis set the ionization of the valence electrons
$\mathrm{4d}$, $\mathrm{5s}$ or above, and the inner electron $\mathrm{4p}$.
From the total of 37 electrons of the target, we freeze 30 of them in the core
$\mathrm{Ar\,3d^{10}\,4s^2\,^1S_{0}^{e}}$,
and the remaining 7 outer electrons, we distribute them in the following set 
of 34 configurations:


\begin{tabular}{ll}
   Even: & $\mathrm{4p^6\,\{4d, 5s, 5d, 6s, 6d\}}$,  \\
         & $\mathrm{4p^5\,5s\,\{4f, 5p, 5f, 6p \}}$, \\
         & $\mathrm{4p^5\,5p\,\{4d, 5d, 6s \}}$,     \\ 
         & $\mathrm{4p^5\,6s\,6p}$,                  \\
         & $\mathrm{4p^4\,\{4d^3, 5s^2\,6s \}}$.     \\
   Odd:  & $\mathrm{4p^6\,\{4f, 5p, 5f, 6p\}}$,      \\
         &  $\mathrm{4p^5\,\{4d^2, 4f^2, 5s^2, 5p^2, 5d^2, 5f^2, 6s^2, 6p^2, 6d^2 \}}$, \\
         & $\mathrm{4p^5\,5s\,\{4d, 5d, 6s \}}$,     \\
         & $\mathrm{4p^5\,5p\,\{5f, 6p \}}$,         \\
         & $\mathrm{4p^4\,4f^3}$.                    \\
\end{tabular}

With this configuration set, we obtain a total of $1\,615$ relativistic levels,
16 of them are spectroscopic, and the rest are 
energetically above the ionization limit and are embedded in the continuum.

\begin{table*}
\caption{Excitation energies of the 16 spectroscopic levels included 
         in the CI expansion of $\mathrm{Sr^{+}}$.}
\label{tab:enerSr1}
\begin{tabular}{rllrlrrrrr}
\hline
 $  i$ & Configuration       & Term               & $J$ & parity & GRASP      & ($\%$) & NIST      \\
\hline
 $  1$ & $\mathrm{4p^6\,5s}$ & $\mathrm{^2S}$     & $1/2$ & even & $    0.00$ & $  -  $ & $    0.0 $ \\
 $  2$ & $\mathrm{4p^6\,4d}$ & $\mathrm{^2D}$     & $3/2$ & even & $14019.88$ & $ -3.7$ & $14555.90$ \\
 $  3$ &                     &                    & $5/2$ & even & $14187.46$ & $ -4.4$ & $14836.24$ \\
 $  4$ & $\mathrm{4p^6\,5p}$ & $\mathrm{^2P^{o}}$ & $1/2$ & odd  & $23459.14$ & $ -1.1$ & $23715.19$ \\
 $  5$ &                     &                    & $3/2$ & odd  & $24232.04$ & $ -1.2$ & $24516.65$ \\
 $  6$ & $\mathrm{4p^6\,6s}$ & $\mathrm{^2S}$     & $1/2$ & even & $48354.77$ & $  1.3$ & $47736.53$ \\
 $  7$ & $\mathrm{4p^6\,5d}$ & $\mathrm{^2D}$     & $3/2$ & even & $53976.98$ & $  1.3$ & $53286.31$ \\
 $  8$ &                     &                    & $5/2$ & even & $54049.72$ & $  1.3$ & $53372.97$ \\
 $  9$ & $\mathrm{4p^6\,6p}$ & $\mathrm{^2P^{o}}$ & $1/2$ & odd  & $55677.25$ & $ -0.2$ & $55769.7 $ \\
 $ 10$ &                     &                    & $3/2$ & odd  & $55943.95$ & $ -0.2$ & $56057.9 $ \\
 $ 11$ & $\mathrm{4p^6\,4f}$ & $\mathrm{^2F^{o}}$ & $7/2$ & odd  & $63578.02$ & $  4.2$ & $60990.04$ \\
 $ 12$ &                     &                    & $5/2$ & odd  & $63579.81$ & $  4.2$ & $60991.34$ \\
 $ 13$ & $\mathrm{4p^6\,6d}$ & $\mathrm{^2D}$     & $3/2$ & even & $78267.46$ & $ 15.9$ & $67522.87$ \\
 $ 14$ &                     &                    & $5/2$ & even & $78301.60$ & $ 15.9$ & $67563.15$ \\
 $ 15$ & $\mathrm{4p^6\,5f}$ & $\mathrm{^2F^{o}}$ & $5/2$ & odd  & $80150.76$ & $ 12.8$ & $71065.80$ \\
 $ 16$ &                     &                    & $7/2$ & odd  & $80185.01$ & $ 12.8$ & $71065.80$ \\
 \hline
\end{tabular}

\flushleft{Key: $i$: level index;
Conf: dominant electron configuration;
Term: dominant LS term;
J: level angular momentum;
GRASP: present GRASP calculation; 
NIST: recommended value from NIST data base~\citep{nist2018};
$\%$: deviation respect the recommended values of NIST, in percentage.
All energies in $\mathrm{cm}^{-1}$.
}

\end{table*}

\begin{table*}
\caption{Excitation energies of the 16 spectroscopic levels included 
         in the CI expansion of $\mathrm{Y^{2+}}$.}
\label{tab:enerY2}
\begin{tabular}{rllrlrrrrr}
\hline
 $  i$ & Configuration       & Term               & $J$ & parity & GRASP       & ($\%$)  & NIST       \\
\hline
 $  1$ & $\mathrm{4p^6\,4d}$ & $\mathrm{^2D}$     & $3/2$ & even & $     0.00$ & $   - $ & $     0.0 $ \\
 $  2$ &                     &                    & $5/2$ & even & $   620.63$ & $-14.3$ & $   724.15$ \\
 $  3$ & $\mathrm{4p^6\,5s}$ & $\mathrm{^2S}$     & $1/2$ & even & $  9263.19$ & $ 24.1$ & $  7467.1 $ \\
 $  4$ & $\mathrm{4p^6\,5p}$ & $\mathrm{^2P^{o}}$ & $1/2$ & odd  & $ 43563.58$ & $  5.2$ & $ 41401.46$ \\
 $  5$ &                     &                    & $3/2$ & odd  & $ 45074.35$ & $  4.9$ & $ 42954.87$ \\
 $  6$ & $\mathrm{4p^6\,6s}$ & $\mathrm{^2S}$     & $1/2$ & even & $ 89104.89$ & $  2.8$ & $ 86717.59$ \\
 $  7$ & $\mathrm{4p^6\,5d}$ & $\mathrm{^2D}$     & $3/2$ & even & $ 91166.57$ & $  3.2$ & $ 88379.61$ \\
 $  8$ &                     &                    & $5/2$ & even & $ 91357.63$ & $  3.1$ & $ 88578.29$ \\
 $  9$ & $\mathrm{4p^6\,6p}$ & $\mathrm{^2P^{o}}$ & $1/2$ & odd  & $100977.19$ & $  1.6$ & $ 99345.62$ \\
 $ 10$ &                     &                    & $3/2$ & odd  & $101547.33$ & $  1.6$ & $ 99943.71$ \\
 $ 11$ & $\mathrm{4p^6\,4f}$ & $\mathrm{^2F^{o}}$ & $7/2$ & odd  & $107933.91$ & $  6.8$ & $101088.23$ \\
 $ 12$ &                     &                    & $5/2$ & odd  & $107927.27$ & $  6.8$ & $101091.42$ \\
 $ 13$ & $\mathrm{4p^6\,6d}$ & $\mathrm{^2D}$     & $3/2$ & even & $128388.86$ & $  7.9$ & $118936.91$ \\
 $ 14$ &                     &                    & $5/2$ & even & $128489.22$ & $  7.9$ & $119029.3 $ \\
 $ 15$ & $\mathrm{4p^6\,5f}$ & $\mathrm{^2F^{o}}$ & $5/2$ & odd  & $131666.08$ & $  6.9$ & $123192.92$ \\
 $ 16$ &                     &                    & $7/2$ & odd  & $131728.70$ & $  6.1$ & $124193.02$ \\
 \hline
\end{tabular}

\flushleft{Key: $i$: level index;
Conf: dominant electron configuration;
Term: dominant LS term;
J: level angular momentum;
GRASP: present GRASP calculation; 
NIST: recommended value from NIST data base~\citep{nist2018};
$\%$: deviation respect the recommended values of NIST, in percentage.
All energies in $\mathrm{cm}^{-1}$.
}

\end{table*}

\begin{table*}
\caption{Excitation energies of the 16 spectroscopic levels included 
         in the CI expansion of $\mathrm{Zr^{3+}}$.}
\label{tab:enerZr3}
\begin{tabular}{rllrlrrrrr}
\hline
 $  i$ & Configuration       & Term               &   $J$ & parity &     GRASP & ($\%$) & NIST       \\
\hline
 $  1$ & $\mathrm{4p^6\,4d}$ & $\mathrm{^2D}$     & $3/2$ & even & $     0.00$ & $  - $ & $     0.0 $ \\
 $  2$ &                     &                    & $5/2$ & even & $  1173.88$ & $-6.1$ & $  1250.7 $ \\
 $  3$ & $\mathrm{4p^6\,5s}$ & $\mathrm{^2S}$     & $1/2$ & even & $ 42045.60$ & $ 9.9$ & $ 38258.35$ \\
 $  4$ & $\mathrm{4p^6\,5p}$ & $\mathrm{^2P^{o}}$ & $1/2$ & odd  & $ 86690.98$ & $ 5.8$ & $ 81976.5 $ \\
 $  5$ &                     &                    & $3/2$ & odd  & $ 89123.42$ & $ 5.5$ & $ 84461.35$ \\
 $  6$ & $\mathrm{4p^6\,5d}$ & $\mathrm{^2D}$     & $3/2$ & even & $151566.98$ & $ 3.4$ & $146652.4 $ \\
 $  7$ &                     &                    & $5/2$ & even & $151915.60$ & $ 3.3$ & $147002.46$ \\
 $  8$ & $\mathrm{4p^6\,6s}$ & $\mathrm{^2S}$     & $1/2$ & even & $157000.96$ & $ 2.9$ & $152513.  $ \\
 $  9$ & $\mathrm{4p^6\,4f}$ & $\mathrm{^2F^{o}}$ & $5/2$ & odd  & $168684.56$ & $ 6.1$ & $159066.75$ \\
 $ 10$ &                     &                    & $7/2$ & odd  & $168740.83$ & $ 6.1$ & $159086.91$ \\
 $ 11$ & $\mathrm{4p^6\,6p}$ & $\mathrm{^2P^{o}}$ & $1/2$ & odd  & $173533.95$ & $ 2.2$ & $169809.71$ \\
 $ 12$ &                     &                    & $3/2$ & odd  & $174507.06$ & $ 2.2$ & $170815.11$ \\
 $ 13$ & $\mathrm{4p^6\,6d}$ & $\mathrm{^2D}$     & $3/2$ & even & $207260.25$ & $ 4.8$ & $197765.1 $ \\
 $ 14$ &                     &                    & $5/2$ & even & $207443.08$ & $ 4.8$ & $197930.43$ \\
 $ 15$ & $\mathrm{4p^6\,5f}$ & $\mathrm{^2F^{o}}$ & $5/2$ & odd  & $207759.05$ & $ 3.3$ & $201114.14$ \\
 $ 16$ &                     &                    & $7/2$ & odd  & $207923.99$ & $ 3.4$ & $201162.65$ \\
 \hline
\end{tabular}

\flushleft{Key: $i$: level index;
Conf: dominant electron configuration;
Term: dominant LS term;
J: level angular momentum;
GRASP: present GRASP calculation; 
NIST: recommended value from NIST data base~\citep{nist2018};
$\%$: deviation respect the recommended values of NIST, in percentage.
All energies in $\mathrm{cm}^{-1}$.
}

\end{table*}

In tables \ref{tab:enerSr1} -- \ref{tab:enerZr3} we compare the calculated
theoretical excitation energies of the spectroscopic levels of $\mathrm{Sr}^{+}$,
$\mathrm{Y}^{2+}$ and $\mathrm{Zr}^{3+}$ with the observed values tabulated in
the NIST atomic spectra data base \citep{nist2018}.
As expected, we achieve the best agreement with the recommended data for the
higher charged ions.
For $\mathrm{Sr}^{+}$, we reach an agreement below $5\%$ for most of the levels.
The most diffuse orbitals, 6d and 5f, have a larger deviation, these orbitals were
very difficult to converge in a single-ionized atom.
$\mathrm{Sr}^{+}$ is the only of the three systems with the energy of the 5s orbital  
below the 4d, in the other two ions, the ground configuration corresponds to the 4d.
For $\mathrm{Y}^{2+}$, the deviations are also the order of the $5\%$ except for the
most diffuse orbitals 4f, 5f and 6d.
In this case, the 4d is the ground configuration and the 5s is very close in energy, 
with only $7467.1\,\mathrm{cm^{-1}}$ of energy difference.
We conclude that the relative error is small, and artificially increased for 
these levels.
For $\mathrm{Zr}^{3+}$, the deviation remains below $5\%$.
As before, the relative error is enhanced for the 4d and 5s levels, 
but remains small and is less noticable than in the previous case.
We also note the opposite effect for the levels 6d and 5f, 
in these cases the deviation is larger than for the other orbitals, 
but the relative error is artificially decreased because the absolute numbers 
are large.

\begin{table*}
\caption{Calculated oscillator strengths for transitions in the optical 
         of the ion $\mathrm{Sr^{+}}$.}
\label{tab:radSr1}
\begin{tabular}{rrrrccccccc}
  \hline
  $i$ & $j$ & Low lev & Up lev & $\lambda$ & $\lambda$ air & $\lambda_{GR}$ & GRASP L & GRASP V & $\lambda$ corr & CIV3 \\
  \hline
  $1$ & $5$ & $\mathrm{5s\,^2S_{1/2}}$ & $\mathrm{5p\,^2P_{3/2}^{o}}$ &  $4078.86$ & $4077.714$ & $4126.8$ & $1.44$  & $1.44$  & $1.46$  & $1.39$ \\
  $1$ & $4$ & $\mathrm{5s\,^2S_{1/2}}$ & $\mathrm{5p\,^2P_{1/2}^{o}}$ &  $4216.71$ & $4215.524$ & $4262.7$ & $0.698$ & $0.698$ & $0.706$ & $0.668$ \\
  \hline
\end{tabular}

\flushleft{Key: $i$: lower level index; $j$: upper level index;
$\lambda$: transition wavelength in \AA, observed in vacuum;
$\lambda$ air: transition wavelength in \AA, observed in air;
$\lambda_{GR}$: transition wavelength in \AA, calculated with GRASP;
GRASP~L: $gf$ value calculated with GRASP in length gauge;
GRASP~V: idem in velocity gauge;
$\lambda$ corr: $gf$ in length gauge corrected the wavelength with the observed value; 
CIV3: $gf$ value calculated with CIV3 in reference \citep{naslim2011}.
}

\end{table*}

\begin{table*}
\caption{Calculated oscillator strengths for transitions in the optical 
         of the ion $\mathrm{Zr^{3+}}$.}
\label{tab:radZr3}
\begin{tabular}{rrrrcccccccc}
  \hline
  $i$ & $j$ & Low lev & Up lev & $\lambda$ & $\lambda$ air & $\lambda_{GR}$ & GRASP L & GRASP V & $\lambda$ corr & CIV3 \\
  \hline
  $6$ & $12$ & $\mathrm{5d\,^2D_{3/2}}$ & $\mathrm{6p\,^2P_{3/2}^{o}}$ &  $4138.6$ & $4137.435$ & $4359.2$ & $0.243$ & $0.226$ & $0.256$ & $0.237$ \\
  $7$ & $12$ & $\mathrm{5d\,^2D_{5/2}}$ & $\mathrm{6p\,^2P_{3/2}^{o}}$ &  $4199.4$ & $4198.265$ & $4426.5$ & $2.19$  & $2.04$  & $2.31$  & $2.10$ \\
  $6$ & $11$ & $\mathrm{5d\,^2D_{3/2}}$ & $\mathrm{6p\,^2P_{1/2}^{o}}$ &  $4318.3$ & $4317.081$ & $4552.3$ & $1.20$  & $1.13$  & $1.27$  & $1.17$ \\
  \hline
\end{tabular}

\flushleft{Key: $i$: lower level index; $j$: upper level index;
$\lambda$: transition wavelength in \AA, observed in vacuum;
$\lambda$ air: transition wavelength in \AA, observed in air;
$\lambda_{GR}$: transition wavelength in \AA, calculated with GRASP;
GRASP~L: $gf$ value calculated with GRASP in length gauge;
GRASP~V: idem in velocity gauge;
$\lambda$ corr: $gf$ in length gauge corrected the wavelength with the observed value; 
CIV3: $gf$ value calculated with CIV3 in reference \citep{naslim2011}.
}

\end{table*}

For a finer check of the quality of the atomic structure, we compare the
radiative transition coefficients, calculated with GRASP in the length and
velocity gauges.
The ratio between the values calculated with the different gauges should be 
unitary, fundamentally in the most intense E1 transitions.
In tables \ref{tab:radSr1}, \ref{tab:radZr3} we show the values of the 
oscillator strengths $gf$ of the transitions present in the optical spectrum 
of \mbox{LS\,IV-14\degr 116}.
We compare the $gf$ values in length and velocity gauges, together with the 
previous calculations made by \citet{naslim2011} using the CIV3 
code \citep{hibbert1975}.
The ratio between length and velocity gauges for the intense E1 transitions
is of the order $1.0$ for the $\mathrm{Sr^{+}}$ and $0.93$ for $\mathrm{Zr^{3+}}$,
and the deviation from the previously calculated values with CIV3 is less than
the $10\%$.
From a simulation perspective, for example as input of {\sc sterne} and {\sc spectrum}, 
it is better to employ modified oscillator strengths that have incorporated the NIST 
observed energy values for the atomic structure.
In tables we also show this corrected values of $gf$, 
its variation is quite small, below the $10\%$.

For the online material, we provide the complete table for the radiative transition 
coefficients E1, M1, E2 and M2 obtained with GRASP between all the spectroscopic levels 
of the targets $\mathrm{Sr}^{+}$, $\mathrm{Y}^{2+}$ and $\mathrm{Zr}^{3+}$.

\section{Scattering}
\label{sec:scattering}

We use a fully relativistic Dirac atomic R-matrix code 
(DARC) \citep{ait-tahar1996,norrington1981,norrington1987}
to calculate the photoionization cross sections 
of $\mathrm{Sr}$, $\mathrm{Y^{+}}$ and $\mathrm{Zr^{2+}}$
initially in their ground or excited levels.
In the inner region, we employ the dipole version of the DARC package, 
including the recent upgrades to optimise the usage of memory and 
processor load balancing using the MPI protocol 
\cite{ballance2004,smyth2019b}.

In the outer region, we use the community code 
PSTGBF0DAMP (Gorczyca and Badnell, unpublished).
PSTGBF0DAMP calculates bound-continuum transitions in the $N+1$-electron atom, 
produced by photon absorption.
In consequence, apart from the Hamiltonian matrixes for each $J \pi$ symmetry, 
we need to calculate with DARC the dipole matrixes between the possible pairs
of E1-connected symmetries.
We have also to calculate the bound states of each symmetry of the 
$N+1$-electron system.
For that goal, we use the STGB program \cite{seaton1982,berrington1987}. 
STGB was recently upgraded by our group, we implemented a CUDA-FORTRAN environment 
to take advantage of the modern GPU-based architectures.
This upgrade reduced the computation time up to a factor 20 in some cases.
Finally, we use the bound-bound program STGBB to calculate the
electric dipole photoexcitation transition coefficients of 
the $N+1$-electron atom.

To run PSTGBF0DAMP, we split the photon-energy grid in two different sets.
Firstly, a fine mesh in order to map the narrow resonances at low energies.
This set contains at least the electron final energies up to the last spectroscopic 
level included in the CC expansion.
In this section, we fix the energy mesh to a fine value of $10^{-5} z^2 \Ry$, 
with $z$ the final ion charge.
The second set is above the final level included within the close-coupling expansion, 
above which we expect no fine resonance structure,
so we can set a coarser mesh. 
This section extends from the end of the previous fine mesh up to approximately 
two times the second ionization threshold, 
this upper limit is far enough to model opacities of photoionized plasmas at 
temperatures of the order of $40\,000 \kelvin$.
We fix this coarse energy mesh to $10^{-3} z^2 \Ry$.
Above the limit of double-ionization, additional processes should be taken into 
account, such as Auger autoionization.
This processes can affect the resonance structure of the cross sections, 
but not to the background.
With the present method, we can calculate with good accuracy photoionization cross
sections for photon energy up to at least $6 \Ry$ in the $\mathrm{Sr}$ atom,
$8 \Ry$ in the $\mathrm{Y}^{+}$ and $10 \Ry$ in the $\mathrm{Zr}^{2+}$, 
which is more than enough to model the opacity in the optical band of the spectrum 
of \mbox{LS\,IV-14\degr 116} or other hot subdwarf stars.

We include in the close-coupling (CC) expansion enough levels to make sure we
have no pseudoresonances in a wide enough range of photon energies.
For the three ions we include a total of 300 levels in the CC expansion 
of the total of 1615 calculated in the CI calculation.
Then we are sure we have cross sections free of pseudoresonances for 
ejected-electron energies lower than $2.38 \Ry$ for the $\mathrm{Sr}$,
$2.68 \Ry$ for the $\mathrm{Y^{+}}$ and
$3.71 \Ry$ for the $\mathrm{Zr^{2+}}$.
We also review all the calculated cross sections to double check if pseudoresonances
are present at energies above these maxima.
To have a good description of the ionization continuum in the whole range
of calculated photon energies we include a total of
36 continuum functions in the case of $\mathrm{Sr}$,
30 for $\mathrm{Y^{+}}$, and 26 for $\mathrm{Zr^{2+}}$.

We calculate the photoionization of each parent ion initially in its ground
level and excited levels up to approximately the 20th.
All these initial levels are contained in the symmetries with
initial angular momenta $J=0-4$ and both parities.
The maximum number of channels in a symmetry is 
1604 for the $\mathrm{Sr}$, 1608 for the $\mathrm{Y^{+}}$, 
and 1667 for the $\mathrm{Zr^{2+}}$.

The atmosphere of \mbox{LS\,IV-14\degr 116} is a photoionized plasma, 
and the temperature of the chromosphere of the star is of $34\,000 \kelvin$,
equivalent to $0.22 \Ry$.
In this case, photoionization cross sections with photon energies up to the 
second ionization potential of 
$\mathrm{Sr}$, $\mathrm{Y^{+}}$ and $\mathrm{Zr^{2+}}$
constitute a complete enough data set to construct a good opacity model.
These calculated cross sections also fit for the modelling of other subdwarf 
stars SdA and SdO, as all of them have surface temperatures in this range.

\section{Results}
\label{sec:results}

\begin{figure}
   \includegraphics[width=0.95\columnwidth]{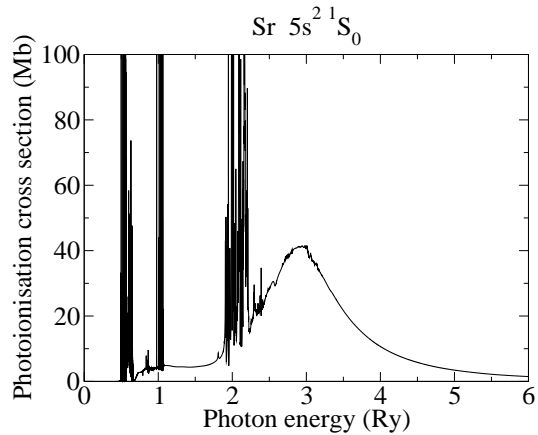}
\caption{Photoionization cross section of $\mathrm{Sr}$
  initially in its ground level versus photon energy.}
\label{fig:piSr0_1S0}
\end{figure}

\begin{figure}
   \includegraphics[width=0.95\columnwidth]{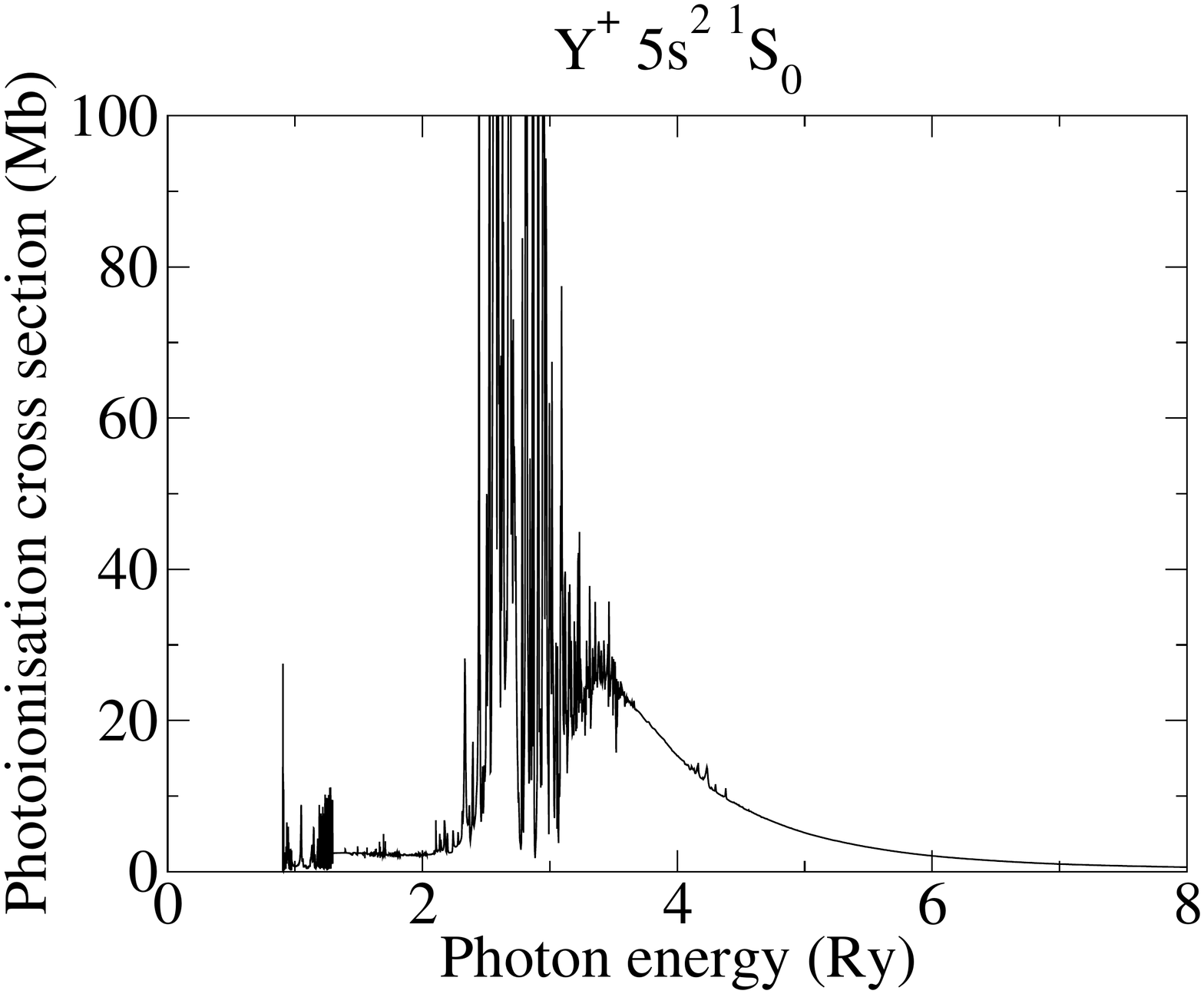}
\caption{Photoionization cross section of $\mathrm{Y^{+}}$
  initially in its ground level versus photon energy.}
\label{fig:piY1_1S0}
\end{figure}

\begin{figure}
   \includegraphics[width=0.95\columnwidth]{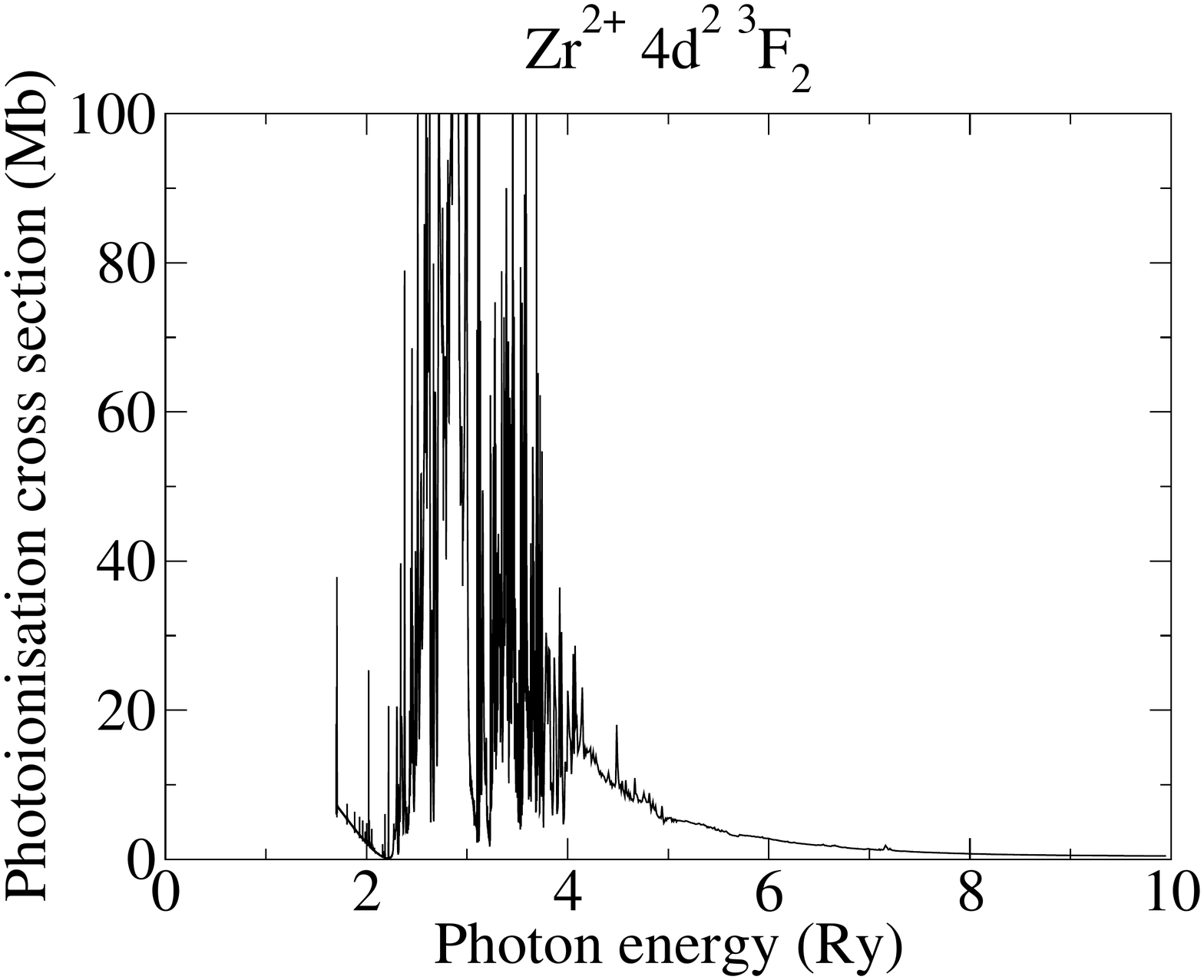}
\caption{Photoionization cross section of $\mathrm{Zr^{2+}}$
  initially in its ground level versus photon energy.}
\label{fig:piZr2_3F2}
\end{figure}

\begin{figure}
   \includegraphics[width=0.95\columnwidth]{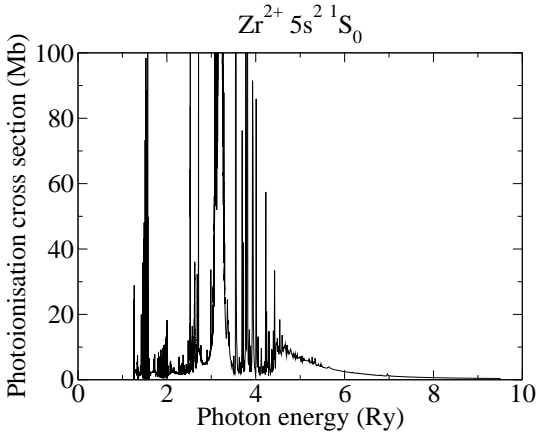}
\caption{Photoionization cross section of $\mathrm{Zr^{2+}}$
  initially in the excited level $\mathrm{5s^2\,^1S_0}$ 
  versus photon energy.}
\label{fig:piZr2_5s2_1S0}
\end{figure}

In figures \ref{fig:piSr0_1S0} -- \ref{fig:piZr2_3F2}, we show the calculated
photoionization cross sections of $\mathrm{Sr}$, $\mathrm{Y^{+}}$ 
and $\mathrm{Zr^{2+}}$ initially in their ground level.
We get the usual structure of a photoionization cross section, 
of resonances superimposed over a background.
The background shows two big increases with posterior decays, 
these big growths correspond to the ionization of an individual orbital, 
the first one of the outer electron, 4d or 5s, and the second, one of the
inner electron 4p.
For the lower-charged ions, $\mathrm{Sr}$ and $\mathrm{Y^{+}}$, 
the ground configuration is the $\mathrm{5s^2}$.
As the charge increases, the relative energy of the 4d orbital reduces, 
and for $\mathrm{Zr^{2+}}$ it positions below the 5s. 
In consequence, the ground configuration of $\mathrm{Zr^{2+}}$ changes
to the $\mathrm{4d^2}$, in opposition to the other two ions.
The photoionization cross section from the ground level of $\mathrm{Zr^{2+}}$
has a slightly different behaviour with respect to the other two ions, 
in particular at photon energies below the second threshold, 
when the ionized electron is the 4d instead of the 5s.
For the ionization from the 4d, the value of the cross section just above the
threshold is not zero.
For higher energies, above the ionization threshold of the inner 4p electron, 
the background of the cross section is similar for the three ions.
For comparison with the ground state of the other two ions,
we show in figure \ref{fig:piZr2_5s2_1S0} the photoionization cross section
of $\mathrm{Zr^{2+}}$ initially in the excited level $\mathrm{5s^2\,^1S_0}$.
In this case, the cross section starts at zero just above the threshold.

For the online material we provide the complete set of photoionization 
cross sections from ground and excited levels of 
$\mathrm{Sr}$, $\mathrm{Y^{+}}$, and $\mathrm{Zr^{2+}}$
in TOP format \citep{opacity1995},
and the complete set of bound-bound photoexcitation transition coefficients 
for all the electric-dipole allowed transitions within the same set
of levels for these three parent ions.
To ensure the files maintain a reasonable downloadable size, order megabytes, 
we convoluted the cross sections to reduce their size to around
one thousand points in each initial level.
The complete photoionization cross sections, without any convolution,
are available for any user under request.
These cross sections can be used for modelling the opacity of 
astrophysical objects, such as stellar atmospheres.

\section{Modelling of stellar atmosphere}
\label{sec:modelling}

\begin{figure}
   \includegraphics[width=0.95\columnwidth]{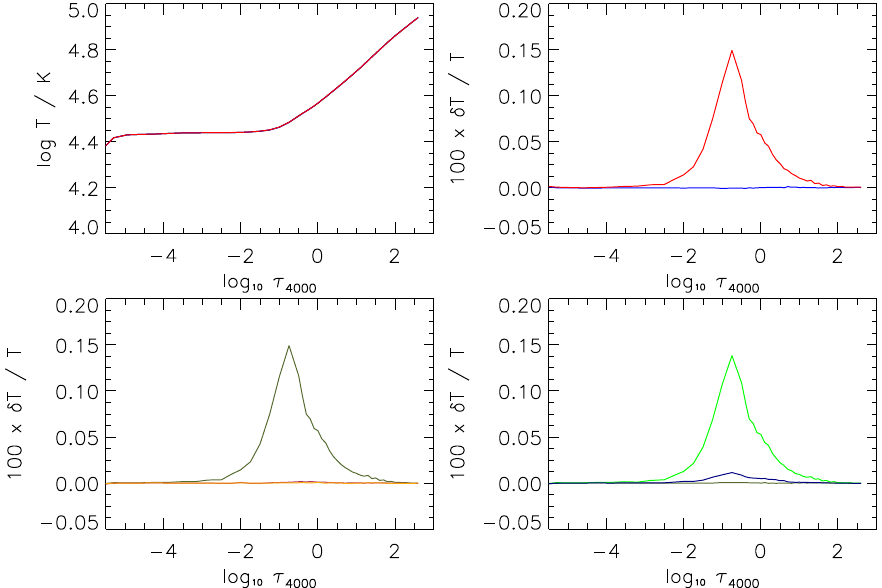}
\caption{The temperature structure of model atmospheres with a 
$20\,000$-fold ($+4.3$ dex) enhancement of one or more of the s-process elements 
strontium, yttrium and germanium are compared with a model having a 
near-solar abundance of the same elements. 
All models have $T_{\mathrm{eff}}=34\,000\,{\mathrm{K}}$, 
$\log g/\,{\mathrm{cm\,s^{-2}}}=5.5$ 
and other properties as shown in table \ref{tab:model}
All panels are plotted as functions of monochromatic optical 
depth $\tau_{4000}$.
Trans-iron ions included are: $\mathrm{Sr^{0}}$, $\mathrm{Sr^{+}}$, $\mathrm{Y^{+}}$, 
$\mathrm{Y^{2+}}$, $\mathrm{Zr^{2+}}$, $\mathrm{Zr^{3+}}$. 
Top left: temperature structure $T/\,(\mathrm{K})$ for both the reference 
model (near solar abundances of Sr, Y and Zr) and the final
model (Sr, Y and Zr abundances increased by $+4.3$ dex);
top right: relative difference ($\delta T/T$ per cent) between models with normal 
and $+4.3$ dex enhanced abundances of Sr, Y and Zr contributing to 
a) photo-ionization only (blue), and 
b) photo-ionisation and photo-excitation combined (red);
bottom left: relative difference between models with and without photo-excitation 
showing the contributions of Sr (orange), Y (maroon) and Z (olive) separately;
bottom right: relative difference between models with and without photo-excitation 
showing the contributions of no photo-excitation (olive), 
$\mathrm{Zr^{2+}}$ (green) and $\mathrm{Zr^{3+}}$ (blue) 
photo-excitation separately.}
\label{fig:tempratio}
\end{figure}

\begin{figure}
   \includegraphics[width=0.95\columnwidth]{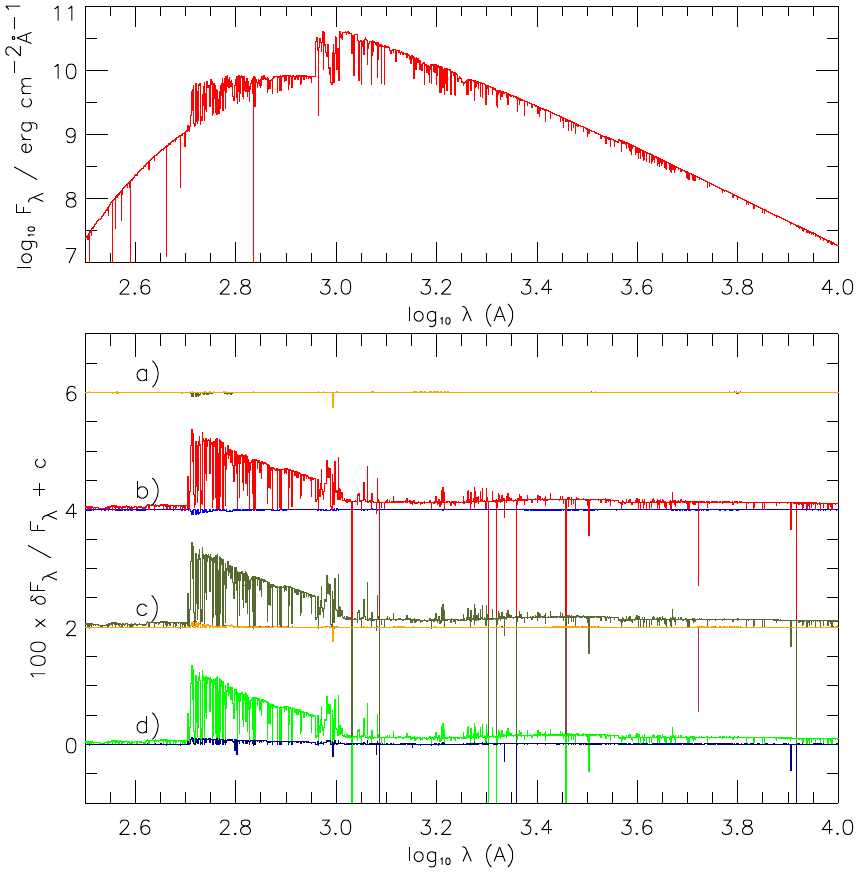}
\caption{The flux emergent from model atmospheres shown in Fig. \ref{fig:tempratio}.
All panels are plotted as functions of wavelength $\log \lambda/\,(\angstrom)$
with $300\,\angstrom < \lambda < 10\,000\,\angstrom$.
Top: The overall flux distribution for the final model including photo-ionisation 
and photo-excitation of Sr, Y and Zr enhanced by $+4.3$ dex; 
bottom: a) (displaced by $+6\%$) relative difference between 
models with Sr (orange), Y (maroon) and Zr (olive) enhanced individually 
by $+4.3$ dex and with no enhancement, 
including photo-ionisation only (all models are essentially indistinguishable); 
b) ($+4\%$) relative difference between models with Sr, Y and Zr enhanced 
collectively by $+4.3$ dex and with no enhancement, and including photo-ionisation 
only (blue) and also photo-excitation (red).
c) ($+2\%$) relative difference between models with and without photo-excitation 
showing the contributions of Sr (orange), Y (maroon) and Zr (olive) separately 
(Sr and Y are indistinguishable);
d) relative difference between models with and without photo-excitation showing 
the contributions of $\mathrm{Zr^{2+}}$ (green) and $\mathrm{Zr^{3+}}$ (blue) 
photo-excitation separately.  
$\mathrm{Zr^{2+}}$ lines provides the major opacity source.}
\label{fig:fluxes}
\end{figure}

\begin{table}
\caption{Model atmosphere parameters for \mbox{LS\,IV-14\degr 116}}
\label{tab:model}
\begin{center}
\begin{tabular}{ccc}
  \hline
  $T_{\mathrm{eff}}/\,{\mathrm{K}}$           & $34\,000$ & \\
  $\log g/\,{\mathrm{cm\,s^{-2}}}$            & $5.5$     & \\
  $v_{\mathrm{turb}}/\,{\mathrm{km\,s^{-1}}}$ & $5$       & \\[1mm]
  $\frac{n_{\mathrm{H}}}{n}$  &  0.64  & \\[1mm]
  $\frac{n_{\mathrm{He}}}{n}$ &  0.32  & \\[1mm]
  $\frac{n_{\mathrm{C}}}{n}$  &  0.014 & \\[1mm]
  $\frac{n_{\mathrm{N}}}{n}$  &  0.014 & \\[1mm]
  $[\mathrm{O}]$      & $-0.8 $  & \\
  $[\mathrm{Si}]$     & $-0.8 $  & \\
  $[\mathrm{Ca}]$     & $-0.25$  & $Z = 17 - 25$ \\
  $[\mathrm{Fe}]$     & $-0.55$  & $Z \geq 26$ \\
  $[\mathrm{Sr}]$     & $-0.25$  $+4.05$ & \\
  $[\mathrm{Y}]$      & $-0.25$  $+4.05$ & \\
  $[\mathrm{Zr}]$     & $-0.25$  $+4.05$ & \\
  \hline
\end{tabular}
\end{center}
\end{table}

Using {\sc sterne},  we calculated models for the atmosphere 
of \mbox{LS\,IV-14\degr 116} assuming base parameters given in 
Table\,\ref{tab:model} for effective temperature $T_{\mathrm{eff}}$, 
surface gravity $g$, microturbulent velocity $v_{\mathrm{turb}}$, 
fractional abundances by number for hydrogen, helium, carbon and nitrogen $n_i$, 
and logarithmic abundances of heavier species relative to solar 
$[\mathrm{X}] \equiv \log \frac{n_{\mathrm{X}}}{n_{\mathrm{X}\odot}}$.  
For Sr, Y and Zr, we used two values, one being scaled to the light metals 
Ca -- Mn (the reference model), and the second enhanced by a 
factor $20\,000$ ($+4.3$ dex), the latter corresponds approximately to the 
overabundances observed in \mbox{LS\,IV-14\degr 116}.

We construct ten different models in three groups. 
All models include the photoionisation cross-sections calculated here 
for $\mathrm{Sr^{0}}$, $\mathrm{Y^{+}}$ and $\mathrm{Zr^{2+}}$.
The reference model has no enhancement in the abundances of these elements.
There are four models which include abundances of Sr, Y, and Zr enhanced 
by $+4.3$ dex, each individually and all collectively.
The third group of models includes a revised linelist with the calculated 
photo excitation transitions of the parent ions $\mathrm{Sr^{0}}$, 
$\mathrm{Y^{+}}$, $\mathrm{Zr^{2+}}$, and the ionization products
$\mathrm{Sr^{+}}$, $\mathrm{Y^{2+}}$, $\mathrm{Zr^{3+}}$.
Again Sr, Y, and Zr are enhanced by $+4.3$ dex, individually and collectively. 
In addition, two models were computed with the $\mathrm{Zr^{2+}}$ 
and $\mathrm{Zr^{3+}}$ lines included separately.

Figure \ref{fig:tempratio} (top left) shows the temperature structure of model 
atmospheres $T$ versus optical depth $\tau_{4000}$ for different abundances of 
Sr, Y and Zr. 
Remaining panels show the temperature relative difference (per cent) to 
the reference model for various cases.
Results imply that the new photo-ionisation cross section makes negligible 
difference to the total opacity, even with the high abundances adopted.
The photo-excitation transitions make a modest temperature difference in the 
line-forming region ($0.01 < \tau_{4000} < 1$), but only due to a number of 
very strong Zr lines.
This difference corresponds to a temperature increase of up to~$0.15\%$.

The consequences for the overall flux distribution are also modest and are 
illustrated in figure \ref{fig:fluxes}, where the emergent fluxes 
$F_{\lambda}\,(\mathrm{erg\,cm^{-2}\,\angstrom^{-1}})$ (top) 
and differences relative to the reference model (middle and bottom) are shown 
as a function of wavelength $\lambda\,(\angstrom)$.
Again, only the inclusion of the photo-excitation process for Zr ions
significantly disturbs the flux distribution, 
indicating substantial flux blocking by these ions.

\section{Conclusions}
\label{sec:conclusions}

We present high-quality photoionization cross sections of $\mathrm{Sr}$, 
$\mathrm{Y^{+}}$, and $\mathrm{Zr^{2+}}$, initially in their ground and excited 
levels.
We present also the bound-bound photoexcitation transition coefficients for
the same three ions and their ionization products
$\mathrm{Sr^{+}}$, $\mathrm{Y^{2+}}$, and $\mathrm{Zr^{3+}}$.
We calculate all the symmetries with total angular momentum $J=0-4$ and both parities.
With the set of configurations included in the basis set, 
we are able to calculate the cross sections from the initial level up to the 
twentieth excited.

We include a basis set of 34 non-relativistic interacting configurations 
to get the structure of the targets
$\mathrm{Sr^{+}}$, $\mathrm{Y^{2+}}$, and $\mathrm{Zr^{3+}}$.
We get accurate energies and transition coefficients for all the first 
16 spectroscopic levels.

The results of photoionization and photoexcitation are available to be used 
for modelling the opacities of astrophysical objects. 
The transition coefficients, such as the oscillator strengths, to model the 
absorption spectral lines, and to build synthetic spectra.
These synthetic spectra can be compared with observed ones to determine physical
parameters of the astrophysical objects, 
for example chemical abundances of the elements.

In particular, we have used the model atmosphere program {\sc sterne} to test 
the contribution of these new data to the atmospheres of hot subdwarf stars. 
We compare the differences in the models taking into account only the photoionization 
cross sections, and including also the photo-excitation transitions. 
We find that the photo-excitation transitions, from Zr ions in particular, 
contribute significant back warming in the model atmosphere model when this element 
is substantially overabundant, as observed in the heavy-metal 
subdwarf \mbox{LS\,IV-14\degr 116}.

Complete sets of photoionization cross sections and radiative coefficients are
provided in the online material.
Results will be uploaded to the web page of the 
Opacity Project\footnote{\url{http://op-opacity.obspm.fr/opacity/}} 
for public use.
Due to the large size of the files, photoionization cross sections were reduced
using a convolution,
the complete set is available under request.

\section*{Acknowledgments}

Present work has been funded by the STFC through the 
QUB Astronomy Observation and Theory Consolidated Grant ST/P000312/1.
The computation has been performed in the supercomputer Archer, 
property of the Engineering and Physical Science Research Council
under the allocation E585-AMOR.

\bibliographystyle{mnras}
\bibliography{references}


\bsp	
\label{lastpage}
\end{document}